\documentclass[a4paper]{jpconf}
\usepackage{graphicx}
\usepackage{tabularx}
\usepackage{amsmath}

\usepackage{iopams}
\usepackage[sort&compress,numbers,square,comma]{natbib}

\begin{document}
\title{dSECS: Including Variometers in Geomagnetic Field Interpolation}
\author{MJ Heyns$^{1,2}$, SI Lotz$^{1}$ and CT Gaunt$^{2}$}
\address{$^{1}$South African National Space Agency, Space Science Directorate, Hermanus, South Africa\\
$^{2}$Department of Electrical Engineering, University of Cape Town, Cape Town, South Africa}
\ead{mheyns@sansa.org.za}
\begin{abstract}
One of the best tools we have in geophysical modelling is the ability to interpolate the horizontal geomagnetic field at the surface of the Earth. This is especially useful in regions, such as southern Africa, where there is a sparse array of absolute magnetometers available for geomagnetic field measurements. In terms of geomagnetic field interpolation, the spherical elementary current systems (SECS) spatial interpolation scheme has shown to be very successful, and the planar approximation of this method adequate for modelling at mid-latitudes. The SECS interpolation scheme is physics based, making use of the Biot-Savart law and equivalent ionospheric currents to interpolate measured geomagnetic field data. As with most interpolation methods, more data points result in lower error. Therefore, we adapt the SECS method to work with variometers. These instruments measure variations in magnetic field and are more abundant in southern Africa. Merging the two resulting interpolated datasets, the initial absolute geomagnetic field interpolation can be significantly improved. This improved interpolation scheme is not only incredibly useful locally, where a sparse magnetometer array is a challenge, but can also be applied just as effectively in other cases across the globe where there are numerous magnetometers and variometers available.
\end{abstract}
\vspace*{-1cm}
\section{Introduction} \label{sec:intro}
\begin{figure}[ht]
\includegraphics[width=0.47\textwidth]{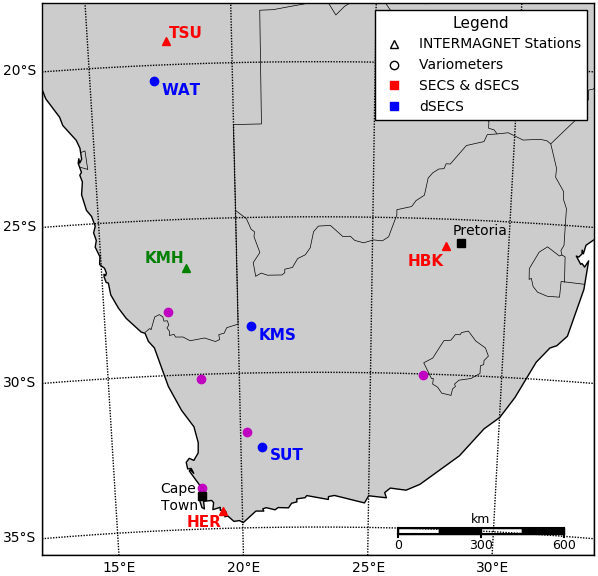}%
\hspace{0.04\textwidth}%
\includegraphics[width=0.54\textwidth]{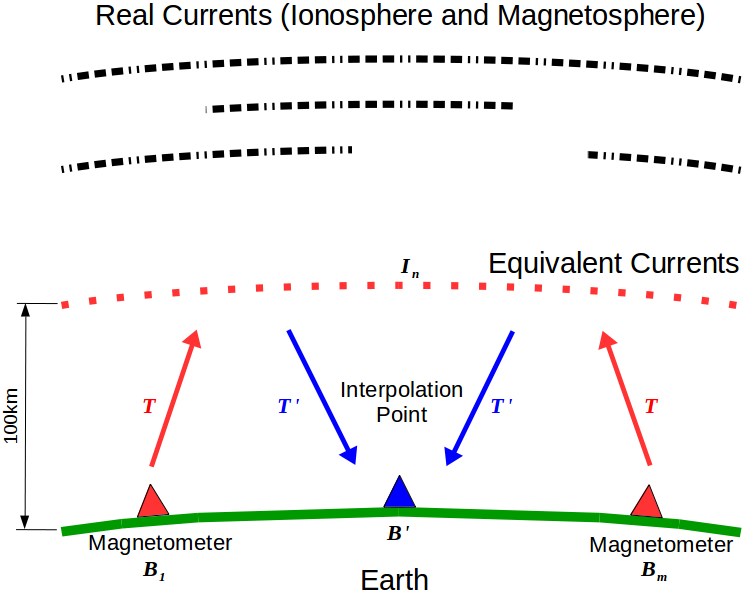}\vspace*{-5pt}
\begin{minipage}[b]{0.47\textwidth}
\caption{\label{fig:map} Map of absolute magnetometers and variometers in southern Africa. Magenta circles represent magnetotelluric stations that can possibly used as variometers as well.}
\end{minipage}
\hspace*{0.04\textwidth}
\begin{minipage}[b]{0.54\textwidth}
\caption{\label{fig:theory} A simple schematic representation of the SECS interpolation scheme and the relevant current systems as described in section \ref{subsec:SECS}.}
\end{minipage}
\vspace*{-22pt}
\end{figure}

In southern Africa, we are lucky enough to have 4 geomagnetic stations which form part of the INTERMAGNET network and measure the absolute geomagnetic field. These stations are located in Hermanus (HER), Hartebeesthoek (HBK), Tsumeb (TSU) and Keetmanshoop (KMH) respectively (see Figure \ref{fig:map}). Although this is a very sparse grid, it is in fact typical and many other regions globally have even fewer stations. Partly what makes these stations so difficult to set-up is the cost involved with the equipment and maintenance needed to measure the baseline geomagnetic field accurately. To get around the sparsity of measurement sites, interpolation has to be used. Much more common worldwide are variometers, which do not measure the baseline, but rather the change in the magnetic field. In southern Africa we have a host of these, for example the pulsation magnetometers at Waterberg (WAT) and Sutherland (SUT) and the magnetotelluric station at Kakamas (KMS). Including the additional information from these denser variometers arrays can improve the geomagnetic field interpolation.
\vspace*{-15pt}\subsection{SECS Interpolation} \label{subsec:SECS}
The spherical elementary current systems (SECS) interpolation technique is a physics based interpolation scheme that adds significant robustness when compared to other purely mathematical interpolation schemes such as Fourier, spherical cap or spherical harmonic interpolation schemes \cite{Amm1999}. From Helmholtz's theorem, any current flowing on a surface can be broken into a curl-free part (which allows current flow in and out of surface) and a divergence-free part (which allows current flow on surface). When considering real world ionospheric and magnetospheric currents, it is the divergence-free part that is typically measured by ground-based magnetometers \cite{Amm1997}. Furthermore, any real system of currents can be approximated by an equivalent current surface at some arbitrary height. Some studies have used multiple current surfaces to separate the external contribution from the induced current contribution. These studies result in estimations of the actual ionospheric and magnetospheric currents themselves \cite{Pulkkinen2003a,Pulkkinen2003b,Marsal2017}. When considering only the horizontal components of the geomagnetic field, it has been shown that a single external equivalent current system suffices \cite{McLay2010}. In addition, SECS interpolation has been shown to be particularly accurate given geomagnetically active days and a sparse grid, making it ideal for a southern African context \cite{McLay2010}.  

Assuming Earth-centred spherical coordinates $(r,\theta,\phi)$, the divergence-free current at a point $\vec{r}$ on the current surface $R_{surf}$, $\theta$ away from a pole at $\vec{r}'$ is, 
\vspace*{-3pt}
\begin{equation}
\label{eqn:main}
\vec{J}_{df}(\vec{r})=\dfrac{I}{4\pi R_{surf}}\cot\left(\dfrac{\theta}{2}\right)\vec{e}_\phi=\dfrac{\iint_S\vec{r}\cdot\vec{\nabla}\times \vec{J}(\vec{r}') dS}{4\pi R_{surf}}\cot\left(\dfrac{\theta}{2}\right)\vec{e}_\phi.
\vspace*{-1pt}
\end{equation}
Assuming cylindrical coordinates ($r=\sqrt{x^2+y^2},\phi,z$)\cite{Viljanen2004}, with a current element of amplitude of $I$ at height $h$, the surface current density would be, 
\vspace*{-3pt}
\begin{equation}
\label{eqn:curden}
\vec{J_{df}}=\dfrac{I}{2\pi r}\vec{e_\phi}.
\vspace*{-1pt}
\end{equation}
There are a number of factors affecting this planar approximation including an small angle assumption and the resolution of stations. Nevertheless, this approximation of the SECS model has been used effectively for geomagnetic field modelling in a southern African context before [\citenum{Bernhardi2008},\citenum{Ngwira2009}]. Assuming $z$ is downwards and a harmonic time dependence (i.e. $e^{i\omega t}$ )\cite{Pulkkinen2003a}, the electric field resulting from the element would be, 
\vspace*{-3pt}
\begin{equation}
\label{eqn:efield}
\vec{E}=-\dfrac{i\omega\mu_0I}{4\pi}\dfrac{\sqrt{r^2+h^2}-h}{r}\vec{e_\phi}.
\vspace*{-1pt}
\end{equation}
The corresponding magnetic field would then be, 
\begin{equation}
\label{eqn:bfield}
\vec{B}=\dfrac{\mu_0I}{4\pi r}\left(\left(1-\dfrac{h}{\sqrt{r^2+h^2}}\right)\vec{e_r}+\left(\dfrac{r}{\sqrt{r^2+h^2}}\right)\vec{e_z}\right).
\vspace*{-1pt}
\end{equation}
Given these governing equations for a single current element at a single point in time, a grid of elementary current elements (say $n$ elements) set-up to cover some defined spatial extent. These elements are collected into a vector $I$. Measured geomagnetic field data from magnetometers (say $m$ stations) is then used to constrain elementary currents, again using the governing equations. As can be expected, $m<n$. These magnetometer stations are collected into a separate vector $B$. To improve accuracy, as much of the Earth's own magnetic field must be subtracted from the measured magnetic field data to ensure that most of the contribution is from the external current systems. Any remaining offset should largely be absorbed by the equivalent currents. A transfer function matrix $T$ relates the elementary currents and the measured geomagnetic field, such that we have the matrix equation $B=T\cdot I$. Since we are only interested in the horizontal field, only the $e_r$ component of the magnetic field is used. This transfer function is only dependent on the spatial relationship between the elementary current and the magnetometer station. The matrix equation calculation is that performed separately for the $x$ (N-S) and $y$ (E-W) components of $I$ and $B$ respectively, and this process is done for each time step. More specifically,
\begin{footnotesize}
\begin{equation}
\label{eqn:secsmat}
\begin{array} {rcl}
\left[ \begin{array}{c} 
B_{x,y:1} \\ 
\vspace{0.15cm} \\ 
\vdots \\ 
\vspace{0.2cm} \\ 
B_{x,y:m} 
\end{array} \right] 
& = & 
\left[ \begin{array}{ccc} 
T_{x,y:11} & \cdots & T_{x,y:1n} \\ & & \\
\vspace{0.15cm} &  & \\
\vdots & \ddots & \vdots \\
\vspace{0.2cm} &  & \\ & & \\
T_{x,y:m1} & \cdots & T_{x,y:mn}
\end{array} \right]
\left[ \begin{array}{c} 
I_1 \\ 
\vspace{0.15cm} \\ \\
\vdots \\ 
\vspace{0.2cm} \\ \\
I_n \end{array} 
\right]
\end{array}
\end{equation}
\end{footnotesize}
where,
\vspace*{-10pt}
\begin{equation}
\label{eqn:transfun}
T_{x,y:ij}=\dfrac{\mu_0}{4 \pi r}\left(1-\dfrac{h}{\sqrt{\smash[b]{r_{ij}^2+h^2}}}\right).
\end{equation}
Since the dimensions of $T$ is $m\times n$ and non-square, $I$ is calculated using the quasi-inverse $T^{-1}$, i.e. $I=T^{-1} \cdot B$. This quasi inverse $T^{-1}$ is obtained from singular value decomposition. Once the vector $I$ is defined by the measured magnetic field, it can be used interpolate the magnetic field to any other point of of interest, i.e. $B'=T'\cdot I$ (see Figure \ref{fig:theory}). The physical consistency and adherence to Maxwell's equations makes the SECS interpolation method incredibly robust.
\vspace*{-10pt}\section{dSECS}
Although variometers are not absolute, they can very accurately measure the change in the magnetic field, i.e. $\Delta B = B(t_i)-B(t_{i-1})$. Since the SECS method is entirely linear in time, $\Delta B$ can be interpolated in the same way as $B$ using the same method ($T$ is purely a spatial constant). The only difference in this case is that $I$ becomes $\Delta I$,
\vspace*{-5pt}
\begin{align}
\label{eqn:dsecs}
B(t_i)-B(t_{i-1})&=T\cdot I(t_i)-T\cdot I(t_{i-1}) \nonumber \\
&=T\cdot\left(I(t_i)-I(t_{i-1})\right) \\
\Delta B &= T\cdot\Delta I. \nonumber
\end{align}
\vspace*{-20pt}
\\With more variometers than absolute magnetometers (magnetometers can act as variometers as well), the confidence in $\Delta B$ interpolation is much higher than that for $B$. $\Delta B$ is also what is typically used for geoelectric studies.
\begin{figure}[hb]
\includegraphics[width=0.465\textwidth]{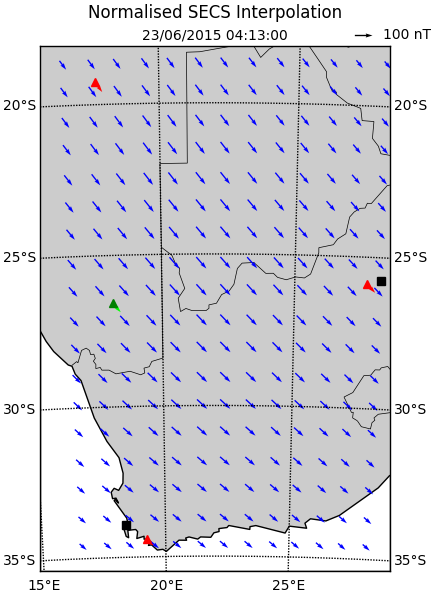}%
\hspace{0.055\textwidth}%
\includegraphics[width=0.48\textwidth]{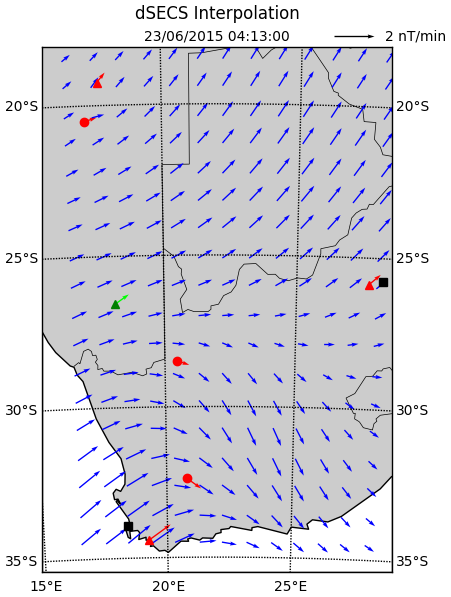}
\vspace*{-20pt}
\caption{\label{fig:typ} Typical results of SECS and dSECS interpolations over southern Africa. Blue vectors indicate the interpolated field and the red vectors indicate the measured field used to interpolate. The green vector is measured data not included in the interpolation and used for validation.}
\vspace*{-15pt}
\end{figure}
\vspace*{-10pt}\section{Merging SECS and dSECS}
Using the greater accuracy in the interpolation of $\Delta B$, we can improve the interpolation of $B$. In order to do this, we consider the two resulting time series of $B$ and $\Delta B$. Let us assume $B$ is of length $N+1$ and hence $\Delta B$ is of length $N$. Given a set of perturbations $\epsilon$, the two resulting time series can be equated,
\vspace*{-10pt}
\begin{align}
\label{eqn:merge}
\Delta B_1 &=(B_2+\epsilon_2)-(B_1+\epsilon_1) \nonumber \\
&\cdots \\
\Delta B_N &=(B_{N+1}+\epsilon_{N+1})-(B_N+\epsilon_N). \nonumber
\end{align}
\vspace*{-20pt}
\\This can then be rewritten in a matrix equation of the form $A\vec{x}=\vec{b}$,
\begin{footnotesize}
\begin{equation}
\label{eqn:mergemat}
\begin{array} {rcl}
\textit{\tiny N}\overset{\xrightarrow[\hphantom{\hspace*{3cm}}]{\textit{\tiny N+1}}}{\left\downarrow\left[ \begin{array}{ccccc} 
-1 & 1 & 0 & \cdots & 0 \\
0 & -1 & 1 & \cdots & 0 \\
\vspace{-0.25cm} & & & & \\
\vdots & \ddots & \ddots & \ddots & \vdots \\
\vspace{-0.25cm} & & & & \\ 
0 & \cdots & 0 & -1 & 1
\end{array} \right]\right.} 
\left[ \begin{array}{c} 
\epsilon_1 \\ 
\epsilon_2 \\ 
\\
\vdots \\ 
\\ 
\epsilon_{N+1} \end{array} 
\right]
& = &
\left[ \begin{array}{c} 
\Delta B_1 + B_1 - B_2 \\ 
\Delta B_2 + B_2 - B_3 \\
\vspace{-0.25cm} \\ 
\vdots \\ 
\vspace{-0.25cm} \\ 
\Delta B_N + B_N - B_{N+1}  
\end{array} \right]. 
\end{array}
\end{equation}
\end{footnotesize}
Although the sparse matrix $A$ is non-square, it has a very well behaved quasi-inverse. This quasi-inverse is again used as before to solve for the perturbations $\vec{x}$, since all the components in $\vec{b}$ are known. Each original interpolated $B_i$ is updated by its corresponding perturbation $\epsilon_i$ for each time step, with the resulting interpolation now also constrained by the more accurate interpolated $\Delta B$.
\vspace*{-10pt}\section{Results}
In order to validate the method an elementary current grid spanning 34.5-18.5$^\circ$S
and 6.5-28.0$^\circ$E was used, which is roughly 1 200km in the East-West direction and 1 700km in the North-South direction. This grid had dimensions 13 x 18 in these respective directions, which in turn relates to a grid spacing of roughly 100km in both directions. In preprocessing the data, the Enhanced Magnetic Model (EMM2017) was used to subtract as much of the Earth's magnetic field as possible to allow the dSECS method to focus on the external contributions. Figure \ref{fig:typ} shows the resulting interpolated magnetic field and the interpolated change in the magnetic field over the this same elementary current grid.

\begin{figure}[ht]
\vspace*{-10pt}
\centering
\includegraphics[width=1.06\textwidth]{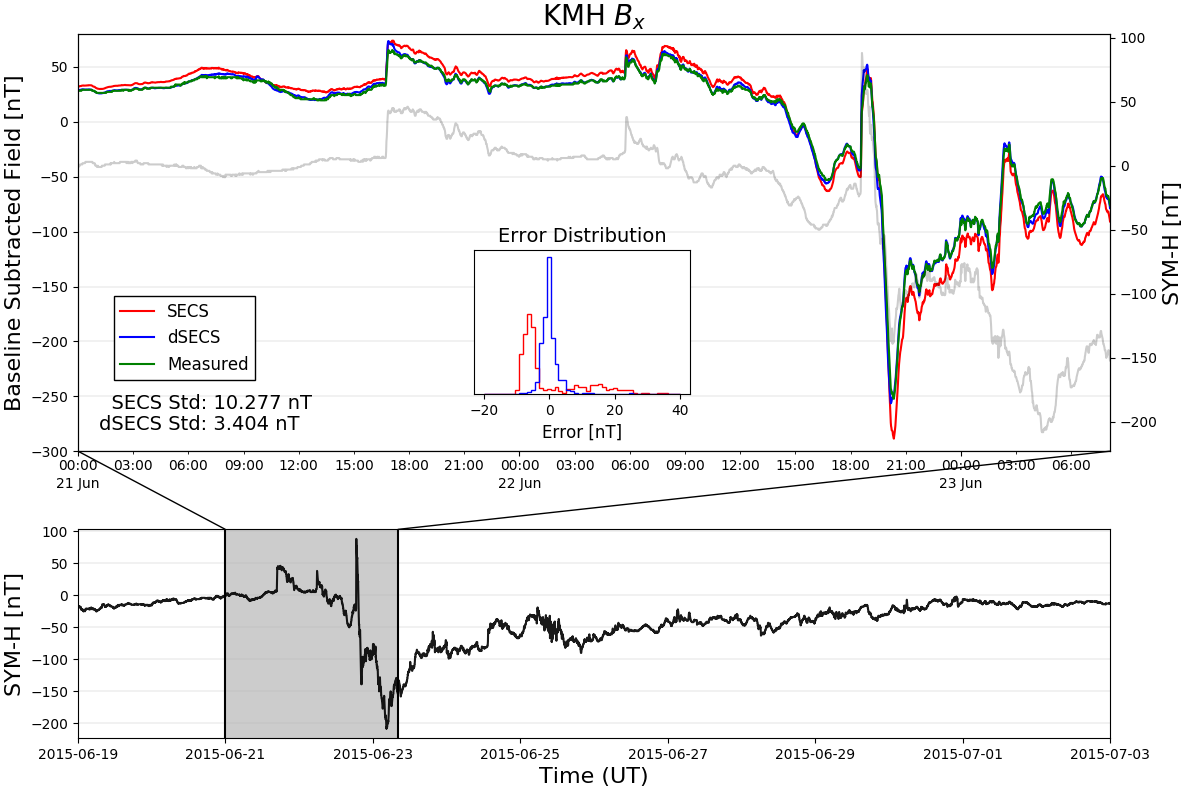}
\vspace*{-20pt}
\caption{\label{fig:storm} The performance of the different interpolation schemes is shown for the $B_x$ component of the geomagnetic field at Keetmanshoop (KMH) during a geomagnetic storm.}
\vspace*{-10pt}
\end{figure}

From the stations described in section \ref{sec:intro}, KMH was used as the validation point for the dSECS method. For this validation, 4 different geomagnetic storms in 2015 were used. There were some data gaps as a result of data availability of all the instruments, but the resulting validation set included quiet time, sudden storm commencement, main phase and recovery phase. Figure \ref{fig:storm} shows the resulting time series for one of these storms for the $B_x$ component. Also shown is the SYM-H index. This is a global index of the ring current and is indicative of the state of the geomagnetic field. For the period shown, there is a distinct sudden storm commencement after a period of quiet time, which is followed by the main phase of the storm. This period has a complex structure with significant fluctuations in the geomagnetic field (large $|\Delta B|$).
\begin{figure}[ht]
\vspace*{-20pt}
\includegraphics[width=0.53\textwidth]{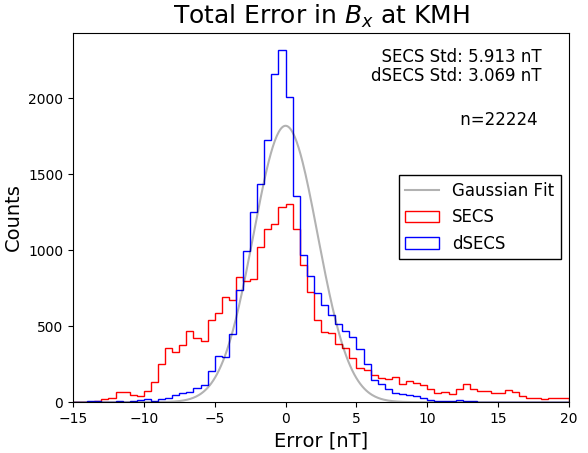}
\includegraphics[width=0.53\textwidth]{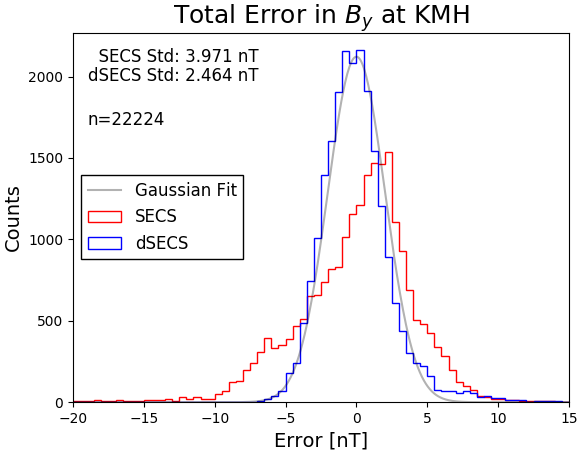}
\vspace{2pt}%
\includegraphics[width=0.53\textwidth]{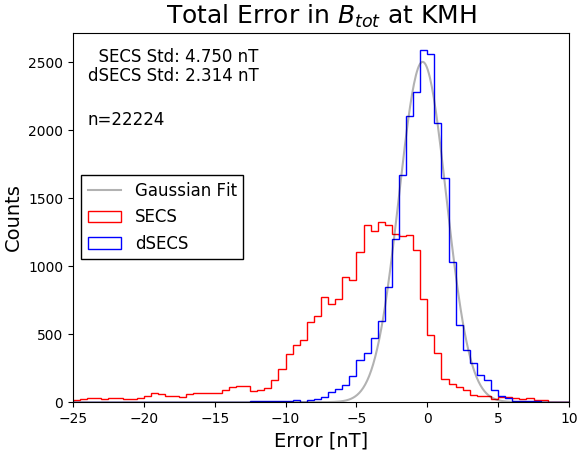}\hspace{0.07\textwidth}%
\begin{minipage}[b]{0.45\textwidth}
\caption{\label{fig:tots} The total error distribution in the baseline subtracted geomagnetic field components for 4 different geomagnetic storms in 2015 are shown, which includes over 22 000 data points. These error distributions are compared to a best-fit Gaussian for the dSECS error distribution.}
\vspace*{1.5cm}
\end{minipage}
\vspace*{-30pt}
\end{figure} 
\vspace*{-10pt}\section{Discussion and Conclusion}
From the resulting modelling and error distributions, it is evident that including the variometers improves geomagnetic field interpolation. When considering the total error distributions (see Figure \ref{fig:tots}), in all magnetic field components the standard deviation is significantly smaller for the dSECS method (between 38-52\% improvement). All dSECS error distributions also tend more towards Gaussian error distributions when compared to the typical SECS method. This suggests that there is less systematic error and more random or sampling error. For the KMH dataset used, there is was a known issue with the decimal point rounding that would a source of such sampling error. It is interesting to note that the $B_y$ component benefits most from the inclusion of variometers. At midlatitudes, this component is typically associated with induction effects which tend to be more localised. Hence the addition of more local variometers is most likely what drives this improvement. Typically, the $B_x$ component is most affected during storm times which may suggest the slightly less Gaussian error distribution. When considering the total horizontal field error though, we again have a roughly Gaussian error distribution which is significantly improved by including variometers.
\\With the dSECS method validated and showing improvement, the method can be used not only for geomagnetic field interpolation but also geoelectric field studies. Geoelectric field studies require $\Delta B$, and hence only variometers are needed. Variometers are much cheaper than absolute magnetometers and more robust, making them feasible for large array implementations. Ultimately, the larger and more dense the array, the greater the interpolation accuracy.
\bibliographystyle{iopart-num}
\vspace*{-8pt}
\footnotesize
\bibliography{proceedings}
\end{document}